\documentclass{aa}
\usepackage{txfonts}
\usepackage{graphicx}
\usepackage{natbib}
\bibpunct{(}{)}{;}{a}{}{,}

\begin{document}
\title{Weak Lensing by High-Redshift Clusters of Galaxies II: Mean Redshift
of the Faint Background Galaxy Population}
\author{D. Clowe\inst{1}\inst{2}\thanks{Visiting Astronomer at the W. M. Keck 
Observatory, jointly operated by the California Institute of Technology and 
the University of California} \and G.A. Luppino\inst{2} \and N. Kaiser\inst{2}}
\institute{Institut f\"ur Astrophysik und Extraterrestrische Forschung der 
Universit\"at Bonn, Auf dem H\"ugel 71, 53121 Bonn, Germany \and
Institute for Astronomy, University of Hawaii,
2680 Woodlawn Drive, Honolulu, HI 96822}
\offprints{D. Clowe, \email{clowe@astro.uni-bonn.de}}
\date{Received 02 May 2003 / Accepted 21 July 2003}

\abstract{
We use weak lensing shear measurements of six $z>0.5$ clusters of galaxies
to derive the mean lensing redshift of the background galaxies used to
measure the shear.  Five of these clusters are compared to X-ray
mass models and {verify} a mean lensing redshift for a $23<R<26.3$,
$R-I<0.9$ background galaxy population in good agreement with photometric
redshift surveys of the HDF-S.  The lensing strength of the six clusters is
also analyzed as a function of the magnitude of the background galaxies,
and an increase in shear with increasing magnitude is detected at
moderate significance.  The change in the strength of the shear is presumed
to be caused by an increase in the mean redshift of the background galaxies
with increasing magnitude, and the degree of change detected is also in
agreement with those in photometric redshift surveys of the HDF-S.
}

\authorrunning{D. Clowe et al.}
\titlerunning{Weak Lensing by High-Redshift Clusters of Galaxies II}

\maketitle

\section{Introduction}
In recent years, spectroscopic surveys of faint galaxies on large aperture 
telescopes have been able to measure the redshift distribution of the galaxy
population brighter than $R=24$ \citep[e.g.][]{CO00.1}.  Obtaining spectroscopic
redshifts for a large sample of galaxies to much fainter magnitudes is
not feasible with the current generation of telescopes and spectrographs.
Photometric redshifts \citep[e.g.][]{FE01.1, BO00.1} have been used on deep multi-color fields
to obtain redshift estimates for galaxies to $R\sim 28$.  These redshift
estimates, however, are uncertain as the only tests on the photometric
redshifts are from the brighter spectroscopically measured galaxies.

Weak gravitational lensing, where one measures the mass of a foreground
object by detecting deviations from an isotropic background galaxy
ellipticity distribution, can be used to obtain an independent estimate of
the mean redshift of a galaxy population.  Because the strength of the lensing
signal varies with both the redshift of the background galaxies and the
redshift of the lensing object, comparing the lensing strength of different
populations of objects both within a given field and across different fields
lensed by varying redshift foreground objects can be used to determine
the mean redshift of the galaxy populations.  This was attempted by
\citet{SM94.1} using a set of three clusters at $z=0.26$, 
$0.55$, and $0.89$.  Based primarily on the lack of lensing observed in the 
high redshift cluster, the data resulted in a best fit for a no evolution
model where the majority of the $I<25$ galaxies were at $z<1$.  It was later
determined, however, that the $z=0.89$ cluster used had a very low X-ray
luminosity \citep{CA94.1}.  If the low X-ray luminosity is interpreted
as a low mass, the lack of a weak lensing signal by this cluster would no
longer constrain the faint galaxies to be at low redshift.

A weak lensing signal was detected in the high-redshift cluster \object{MS 1054$-$0321},
at $z=0.826$, by \citet{LU97.1}, which implied that a large 
fraction of the $I\sim25$ galaxies must be at $z>1$.  
With the goals of determining the mass and dynamical state of X-ray
selected, high-redshift clusters of galaxies and determining the mean redshift
of the faint blue galaxy (FBG) population, we have undertaken a survey
of six $z>0.5$ clusters.  We selected as our sample of clusters the five EMSS 
high-redshift clusters (\object{MS $0015.9+1609$} at $z=0.546$, \object{MS $0451.6-0305$} at 
$z=0.550$, \object{MS $1054.4-0321$} at $z=0.826$, \object{MS $1137.5+6625$} at $z=0.782$, and 
\object{MS $2053.7-0449$} at 
$z=0.583$), which were the only $z>0.5$ clusters published from a
serendipitous X-ray survey at the time, and one from the ROSAT North
Ecliptic Pole survey (\object{RXJ $1716.6+6708$} at $z=0.809$) which was discovered
shortly after we began our survey \citep{HE97.1, GI99.1}. 
The weak lensing analysis of the clusters have been published 
\citep{CL98.1, CL00.1}.  In this paper we present the results of 
our attempts to measure
the mean redshift of the FBG population from their weak lensing signal.

In Sect.~2 we present the weak lensing techniques used in our analysis.  Comparison
of the weak lensing signal and X-ray mass estimates is given in Sect.~3.  In
Sect.~4 we present the results of direct comparison of the lensing signal
of various galaxy populations.  Sect.~5 contains our conclusions.
Throughout this paper, unless otherwise stated, we assume an 
$\Omega _{\mathrm{m}}=0.3, \Lambda =0.7$ universe, parameterize our results
in terms of $H_0 = 100 h$ km/s/Mpc and give all errors as 1$\sigma$.

\section{Weak Lensing}
Because the gravitational potential of a cluster of galaxies bends the
trajectories of light rays which pass by it, the observed galaxies behind
the cluster have been deflected away from the cluster center.  In addition,
the galaxies have been sheared in one dimension, which for a circularly
symmetric cluster is tangential to the cluster center.  This shearing
results not only in a change of the ellipticity of the galaxy, as defined
by the second moments of the surface brightness, but also, because the surface
brightness of the galaxy is preserved, in a magnification of the observed
flux.  These effects are discussed in greater detail in \citet{CL00.1}, as 
well as in the reviews by \citet{BA01.1} and \citet{ME99.1}.

\begin{figure}
\centering
\resizebox{\hsize}{!}{\includegraphics{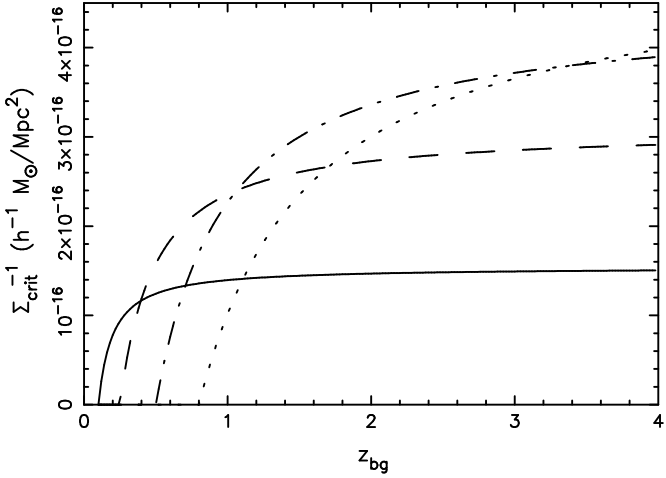}}
\caption{Plotted above are the values of $\Sigma _{\mathrm{crit}}^{-1}$
as a function of background galaxy redshift for four lens redshifts.  The
redshifts of the lenses (0.1, 0.25, 0.5, and 0.8) can be determined from
where $\Sigma _{\mathrm{crit}}^{-1}$ becomes 0.  While the lower redshift
lenses have $\Sigma _{\mathrm{crit}}^{-1}$ only slowing varying over the
expected faint galaxy redshift distribution ($\sim 0.8-2$), the higher
redshift lenses still have $\Sigma _{\mathrm{crit}}^{-1}$ being a strong
function of background galaxy redshift.}
\label{fig1}
\end{figure}

For a single thin lens, such as a cluster of galaxies, the strength of the
lens is expressed in the dimensionless mass surface density $\kappa$, where
\begin{equation}
\kappa = {\Sigma \over \Sigma _{\mathrm{crit}}}.
\end{equation}
$\Sigma $ is the surface density of the cluster, and $\Sigma _{\mathrm{crit}}$ 
is a scaling factor:
\begin{equation}
\Sigma _{\mathrm{crit}}^{-1} = {4 \pi  G \over c^2}{ D_{\mathrm{l}} 
D_{\mathrm{ls}} \over D_{\mathrm{s}}}
\end{equation}
where $D_{\mathrm{s}}$ is the angular distance to the source (background) 
galaxy, $D_{\mathrm{l}}$ is the angular distance to the lens (cluster), 
and $D_{\mathrm{ls}}$ is the angular distance from the lens to the source 
galaxy.  The variation of $\Sigma _{\mathrm{crit}}^{-1}$ with the redshift
of the background galaxies and lens redshift is shown in Fig.~\ref{fig1}.  
As can be
seen, for low redshift $z\le 0.3$ clusters, $\Sigma _{\mathrm{crit}}^{-1}$
quickly rises at redshifts slightly larger than the cluster, but by
$z_\mathrm{bg} \sim 0.8$ changes very little with background galaxy
redshift.  For high redshift ($z\ge 0.5$) clusters, however, this is not
the case, and $\Sigma _{\mathrm{crit}}^{-1}$ continues to increase with
increasing background galaxy redshift throughout the region most background
galaxies are likely to reside ($0.8\le z_{\mathrm{bg}} \le 4$).

\begin{figure*}
\centering
\includegraphics[width=17cm]{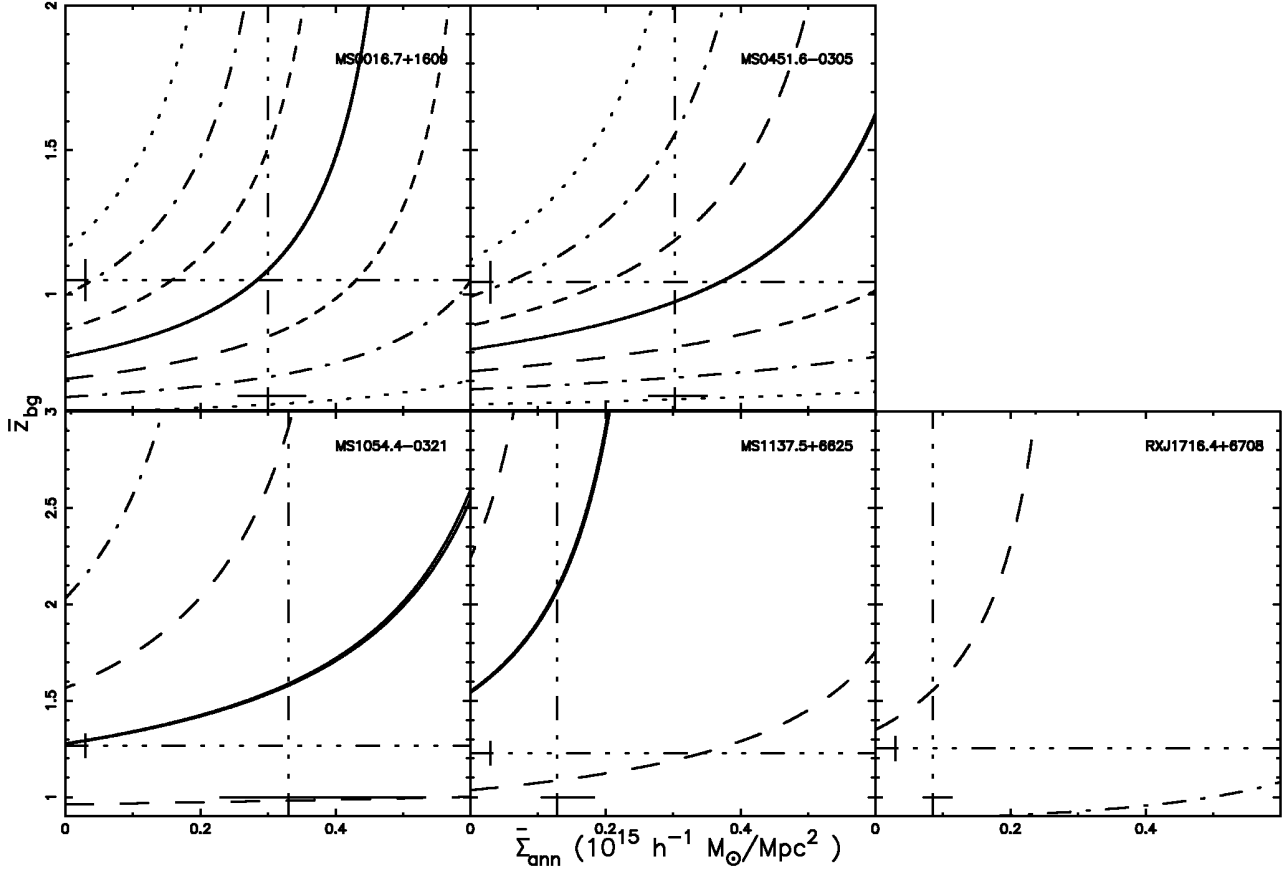}
\caption{Shown above are the best fit values for the mean lensing redshift
of the background galaxy population as a function of 
$\bar{\Sigma}_{\mathrm{ann}}$, which is the mean surface mass density in 
the negative annular region for
$\zeta_\mathrm{c}$ (Eq.~(\ref{eq7})), for the five clusters
with complete X-ray data.  The solid lines are the best fit, and the dashed,
dot-dashed, and dotted lines are the 1, 2, and 3-$\sigma$ deviations from the
best fit respectively.  The vertical lines, with associated error bars, are 
the surface density for the annular region calculated from the best-fit X-ray
$\beta$-model.  The horizontal lines, with associated error bars, are the mean
lensing redshift calculated from the HDF-S photometric redshift catalog
of \citet{FO99.1} using the same magnitude and color cuts as in the observed
data.}
\label{fig2}
\end{figure*}

What is measured from the background galaxies, however, is not $\kappa$, but
the reduced shear $g$, which is related to the gravitational shear $\gamma$
by 
\begin{equation}
g = {\gamma \over 1 - \kappa}.
\end{equation}
When measuring the shear around a circular
aperture, one has the relation that 
\begin{equation}
\langle \gamma (r)\rangle = \bar{\kappa }(<r) - \langle \kappa (r)\rangle
\label{eq4}
\end{equation} 
where $\bar{\kappa }(<r)$ 
is the mean $\kappa $ inside radius $r$ and the $\langle \rangle$ denote
the average over the annulus.  If $\kappa $ has
a small variance along the azimuthally averaged circle,
\begin{equation}
\langle g(r)\rangle  = (\bar{\kappa }(<r) - \langle \kappa (r)\rangle ) 
(1 + \langle \kappa (r) \rangle  + \mathcal{O}(\langle \kappa \rangle ^2)).
\end{equation}
Thus, at large distances from the cluster, where $\kappa \ll 1$, the strength
of the reduced shear signal is effectively linear with 
$\Sigma _{\mathrm{crit}}^{-1}$, and a mean background galaxy lensing redshift
$\bar{z}_\mathrm{bg}$ can be defined for a sample of $N$ galaxies with
\begin{equation}
\Sigma _{\mathrm{crit}}^{-1}(\bar{z}_\mathrm{bg}) = {1\over N} \sum_{i=1}^N
\Sigma _{\mathrm{crit}}^{-1}(z_i).
\end{equation}
By comparing the shears measured from different populations of galaxies
having the same spatial distribution about the lenses, one can then measure
directly a ratio in the mean lensing redshifts for the two populations.

A similar equation can be calculated for the non-weak lensing case,
when $\kappa \sim 1$, however in 
this case the effective mean lensing redshift will be a function of the 
local mass surface density.  Further, due to the competing effects of 
deflection and
magnification of the background galaxies, the redshift distribution of a
magnitude limited sample of the background galaxies will change with
increasing $\kappa$ \citep[e.g.][]{DY01.1}.  As a result, accurately comparing
the mean lensing redshifts of two galaxy populations near the cores of massive
clusters is much harder than at large distances from the cores, and is near
impossible without some pre-existing knowledge of the mass distribution of
the clusters.

As can be seen in Eq.~(\ref{eq4}), in the weak lensing limit a sheet of constant
density across the field can be added to the cluster surface density without
affecting the measured shear.  As a result, one cannot determine $\kappa$ or
$\bar{\kappa}$ at a given point uniquely, but can only determine them to
within an unknown additive constant.  As a result, a useful statistic to
use as a mass estimate is aperture densitometry \citep{FA94.1,CL00.1},
\begin{eqnarray}
\zeta _\mathrm{c}(r_1) & = & \bar{\kappa }(<r_1) - 
\bar{\kappa }(r_2<r<r_\mathrm{max})\nonumber\\ & = &
2\int_{r_1}^{r_2}d\ln r \langle \gamma _\mathrm{T}\rangle + 
{2\over (1-r_2^2/r_\mathrm{max}^2)} \int_{r_2}^{r_\mathrm{max}}d\ln r \langle 
\gamma _\mathrm{T}\rangle,
\label{eq7}
\end{eqnarray}
in which the mean kappa in a constant annular region, 
$\bar{\kappa }(r_2<r<r_\mathrm{max})$, 
is subtracted from the $\bar{\kappa}$ profile of
the cluster.  This statistic is also linear with $\Sigma _{\mathrm{crit}}^{-1}$
and can therefore be combined with a cluster mass measurement at a given radius
from another set of data (X-ray measurements, velocity dispersion, etc)
to determine the mean lensing redshift of the background galaxies.  However,
unless $\bar{\kappa }(r_2<r<r_\mathrm{max})$ can also be determined for the
comparison mass measurement, then there will be a degeneracy between the
assumed $\bar{\kappa }(r_2<r<r_\mathrm{max})$ and the derived
mean lensing redshift.

Further, in the non-weak lensing limit, $\gamma_\mathrm{T}$
in aperture densitometry is replaced by $g_\mathrm{T}$, and the resulting
statistic is no longer measuring $\bar{\kappa }(<r_1) - 
\bar{\kappa }(r_2<r<r_\mathrm{max})$.  The statistic is also no longer linear
with $\Sigma _{\mathrm{crit}}^{-1}$, but can still be used to find a best
fit $\bar{z}_\mathrm{bg}$ by converting a mass profile, which must
cover the same range in $r$ as $\zeta _\mathrm{c}$, to a reduced
shear profile and calculating the resulting $\zeta _\mathrm{c}$
statistic to compare with the measured value.  
If, however, one does have a mass profile determined from an independent data
set, one will typically get a higher signal-to-noise measurement by fitting
the observed reduced shear profile directly with the mass profile converted
to reduced shear profile via the 
$\Sigma _{\mathrm{crit}}^{-1}(\bar{z}_\mathrm{bg})$ fit parameter.  In both
cases, the fitting for the mean lensing redshift can only be done in regions
with a sufficiently low $\kappa $ and $\gamma $ that the magnification and
displacement of the background galaxies do not significantly alter the 
background galaxy redshift distribution.

\section{Comparison with X-ray data}
All six clusters in the study were detected serendipitously in X-ray surveys,
and at the time we created the sample they were the only six $z>0.5$ clusters
discovered in X-ray surveys.  As such, they have been targets of 
extensive studies in X-ray passbands and have published mass models
derived from the measured X-ray luminosities and temperatures.  In Table~\ref{table1}
are the X-ray temperatures and spherical $\beta$-model fits taken from the
literature.  The X-ray temperatures are from ASCA observations and 
$\beta$-model fits from ROSAT observations, except for \object{MS 0451.6$-$0305}, for
which all the values are from Chandra observations, and \object{MS 1054.4$-$0321},
for which the temperature is from Chandra observations.

\begin{table*} 
\centering
{\bf Summary of Cluster Data}\\
\begin{tabular}{lcccccc} \hline
Cluster & z & $\beta$ & $T_X$ & $r_\mathrm{c} $ & 
$\bar{\Sigma}(350 h^{-1} \mathrm{kpc})$ & 
$\zeta _\mathrm{c}(350 h^{-1} \mathrm{kpc})$\\
 & & & (keV) & ($h^{-1}$ kpc) & ($\times 10^{14} h \mathrm{M}_\odot /\mathrm{Mpc}^2$) & \\[0.5ex]\hline
\object{MS 0015.9$+$1609} & $0.547^\mathrm{a}$ & $0.728^{+0.025\mathrm{b}}_{-0.022}$ & $7.55^{+0.72\mathrm{b}}_{-0.58}$ & $182^{+12\mathrm{b}}_{-10}$ & $7.8^{+0.8}_{-0.7}$ & $0.111\pm 0.025$ \\
\object{MS 0451.6$-$0305} & $0.539^\mathrm{c}$ & $0.780^{+0.028\mathrm{d}}_{-0.025}$ & $10.6^{+1.6\mathrm{d}}_{-1.3}$ & $151^{+8\mathrm{d}}_{-7}$ & $12.2^{+1.9}_{-1.5}$ & $0.191\pm 0.025$ \\
\object{MS 1054.4$-$0321} & $0.833^\mathrm{e}$ & $0.96^{+0.48\mathrm{f}}_{-0.22}$ & $10.5^{+3.4\mathrm{g}}_{-2.1}$ & $285^{+118\mathrm{f}}_{-67}$ & $12.5^{+7.4}_{-3.8}$ & $0.220\pm 0.034$\\
\object{MS 1137.5$+$6625} & $0.783^\mathrm{a}$ & $0.70^{+0.27\mathrm{c}}_{-0.09}$ & $5.7^{+1.3\mathrm{c}}_{-0.7}$ & $67^{+33\mathrm{c}}_{-16}$ & $6.3^{+2.8}_{-1.3}$ & $0.156\pm 0.036$\\
\object{MS 2053.7$-$0449} & $0.586^\mathrm{a}$ &     & $8.1^{+3.7\mathrm{h}}_{-2.2}$ & & & $0.076\pm 0.023$\\
\object{RXJ 1716.4$+$6708} & $0.809^\mathrm{i}$ & $0.42^{+0.09\mathrm{i}}_{-0.05}$ & $5.7^{+1.4\mathrm{i}}_{-0.6}$ & $43\pm24^\mathrm{i}$ & $3.8^{+1.3}_{-0.07}$ & $0.138\pm 0.039$\\ \hline
\end{tabular}\\
\caption{
a: \citet{LU95.1}, b: \citet{HU98.1}, c: \citet{YE96.1}, d: \citet{DO03.1},
e: \citet{TR99.1}, f: \citet{NE00.1}, g: \citet{JE01.1}, h: \citet{HE00.1},
i: \citet{GI99.1}.}
\label{table1}
\end{table*}

The standard model used to fit the X-ray data of the clusters is the 
$\beta$-model, for which the mass enclosed in a sphere of radius $r$ is
\begin{equation}
M(r) = 1.13\times 10^{14} \beta T_\mathrm{X}(\mathrm{keV}) r(h^{-1} \mathrm{Mpc}) 
{(r/r_\mathrm{c})^2\over 1+(r/r_\mathrm{c})^2} \mathrm{M}_\odot ,
\end{equation}
where $T_\mathrm{X}$ is the X-ray temperature, $\beta $ and the core radius
$r_\mathrm{c}$ are defined by the gas density $\rho (r) \propto 
[1+(r/r_c)^2]^{-3\beta +1/2}$, and the cluster is assumed to be isothermal
and in hydrostatic equilibrium.  For comparison with weak lensing, the mass
density must be integrated along the line of sight to get the surface mass
density
\begin{eqnarray}
\Sigma (r) & = & 2 \int _{r}^{\infty } {dM(R)/dR\over 4 \pi R^2} 
{R \over \sqrt{R^2 - r^2}} dR \nonumber\\
& = & 2.83\times 10^{13} \beta T_\mathrm{X}(\mathrm{keV}) {r^2 + 2r_\mathrm{c}^2
\over (r^2 + r_\mathrm{c}^2)^{3\over 2}} h \mathrm{M}_\odot /\mathrm{Mpc}^2.
\end{eqnarray}
This can then be integrated over a disk to give the mean surface density
\begin{equation}
\bar{\Sigma }(r) = 5.63\times 10^{13} \beta T_x(keV) {1\over 
\sqrt{r^2+r_\mathrm{c}^2}} h \mathrm{M}_\odot /\mathrm{Mpc}^2.
\end{equation}

Shown in Fig.~\ref{fig2} are the results when the ROSAT mass models are compared to
the weak lensing $\zeta _\mathrm{c}$ statistic at a radius of $350 h^{-1}$
kpc from the cluster center.  The $350 h^{-1}$ kpc radius was arbitrarily
chosen for the comparison as it is the largest radius for which the clusters
have a good signal-to-noise on $\zeta _\mathrm{c}$ but a low enough expected
value of $\kappa $ to have the systematic error in assuming that the measured
value reduced shear $g$ is actually the shear $\gamma$ be insignificant
when compared to the random error on $\zeta _\mathrm{c}$.  Further, $350 h^{-1}$
kpc is also roughly the maximum radius for which all of the clusters have
measured ROSAT X-ray luminosities, and thus not requiring the X-ray mass model
to be extrapolated outside of the region containing measured data.  

In the above comparison, all of the background galaxies were de-magnified
before applying magnitude cuts to the catalog
by the amount $m_\mathrm{demag} = m_\mathrm{orig} - 2.5 \log ((1-\kappa)^2 -
\gamma ^2)$ by assuming that $\kappa (r)= \gamma (r)= g(r)$, and using
the best fit $\gamma ^{-1}$ model to the measured reduced shear over 
a $350 h^{-1} \mathrm{kpc} < r < r_{\mathrm{max}}$ range to calculate
$\gamma (r)$.  This is a first order correction to the magnification of the 
observed background galaxy population, and thus should make the observed
population be on average the same as those observed in blank fields.
{However, because the lensing strength, and thus the magnification, is
a function of the redshift of the galaxies, the average magnification which
we corrected for will be slightly too low for high redshift galaxies and
too high for low redshift galaxies.  Thus, higher redshift galaxies will 
still be slightly magnified and lower redshift galaxies will be slightly 
de-magnified.}  This will result in
a slight overestimation of $\bar{z}_\mathrm{bg}$ by this method, but
from simulations we have determined this systematic error is an order of
magnitude below the random errors in the measurement.  {In future, larger
data-sets, this error could be minimized by binning the data by colors into
groups with similar redshifts.}

\begin{figure}
\centering
\resizebox{\hsize}{!}{\includegraphics{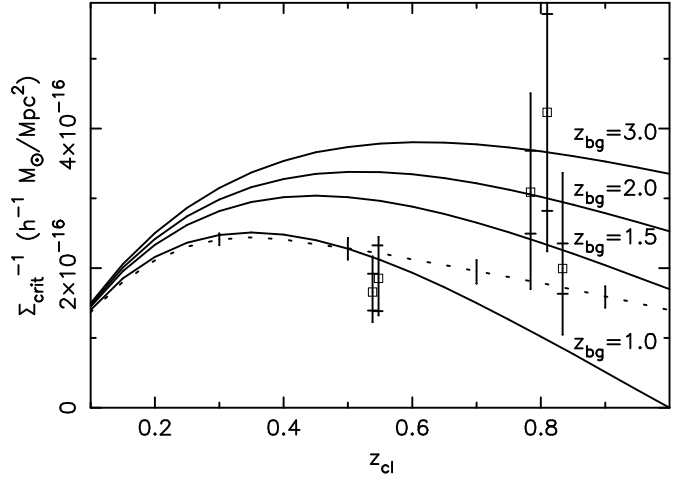}}
\caption{Shown above are the best fit values for $\Sigma _\mathrm{crit}^{-1}$
from comparing the X-ray $\beta$-models with the weak lensing shear profiles
as a function of cluster redshift.  The horizontal bars intersecting the
error bars indicate how much of the error bar is due to the errors in the
weak lensing mass measurement, with the remainder due to the uncertainties in
the X-ray mass measurement.  The four solid curves are the values of
$\Sigma _\mathrm{crit}^{-1}$ for background galaxies at redshifts of
1, 1.5, 2, and 3.  The dashed curve, with error bars, shows the value of
$\Sigma _\mathrm{crit}^{-1}$ from the HDF-S photo-z galaxy catalog.}
\label{fig3}
\end{figure}

As can be seen in Fig.~\ref{fig2}, if the $\beta$-model of the X-ray clusters is
extrapolated to determine the mean $\kappa $ in the annular region subtracted
in $\zeta _\mathrm{c}$, the allowable value for $\bar{z}_{\mathrm{bg}}$ is
in good agreement with that calculated from photometric redshifts of galaxies
with the same magnitude and color range in the HDF-S \citep{FO99.1}.
If one is going to extrapolate the X-ray model over the region containing
the measured reduced shear, however, one will obtain both a better
signal-to-noise and avoid the systematic error of assuming $g$ is $\gamma $
by fitting the reduced shear profile with the $\beta$-model surface mass profile.
In Fig.~\ref{fig3} are the best fit values of $\bar{z}_{\mathrm{bg}}$ when fitting
the shear and mass models over a $300 h^{-1} \mathrm{kpc} < r < r_\mathrm{max}$
range.  The $300 h^{-1}$ kpc inner radius was chosen to avoid the large
changes to the background galaxy redshift distribution which occurs
due to the larger magnifications and displacements of the background galaxies
near the cluster core.

For these fits, the $\beta$-model was converted from $\Sigma $
to $\kappa $ by the fit value $\Sigma _\mathrm{crit}^{-1}$, and used to 
calculate
the reduced shear profile $g(r) = \gamma (r)/(1 - \kappa (r)) = 
(\bar{\kappa}(r) - \kappa (r))/(1 - \kappa (r))$.  The model's $\kappa (r)$ 
and $\gamma (r)$ profiles were then used to calculate and correct for
the average magnification for each background galaxy {as a function of
distance from the cluster center}.  The magnitude corrected
catalog then had the magnitude and color cuts applied to select the catalog
used to measure the reduced shear.  The measured reduced shear was then
compared with the model using a $\chi ^2$ statistic, which was minimized
to find the best fit $\Sigma _\mathrm{crit}^{-1}$.  The resulting 
$\Sigma _\mathrm{crit}^{-1}$ measurements can then be converted to
a $\bar{z}_\mathrm{bg}$ for each cluster.  For a broad background galaxy
redshift distribution, the resulting $\bar{z}_\mathrm{bg}$ is a function
of the lensing cluster redshift due to the change in the 
$\Sigma _\mathrm{crit}^{-1}(z_\mathrm{bg}, z_\mathrm{cl})$ with cluster 
redshift.  The results are in good agreement
with the photometric redshift distribution of faint galaxies from the HDF-S.

{It should be noted that the mean lensing background galaxy redshift is
a function of magnitude, color, size, and surface brightness cuts placed on 
the background galaxy catalog.  Because the images for the five clusters used
in this comparison are similar in exposure times and seeing, the weak lensing
results all use the same background galaxy redshift distribution.  In general,
however, this will not be the case and the mean lensing redshifts as a 
function of cluster redshift shown in Fig.~\ref{fig3} will not be the mean
lensing redshifts of the observations.  For each observation, the mean lensing 
redshift would need to be computed from a redshift catalog by applying the 
same cuts as are used to select the background galaxies.}

\section{Changes in shear strength with magnitude}

As was discussed in Sect.~2, for a high-redshift lens, the strength of the
shear acting on a background galaxy greatly depends on the angular distance of the
background galaxy.  As the ellipticity induced in the galaxy by the
weak lensing shear is smaller than the typical intrinsic ellipticity of the
galaxy, one cannot use this to determine angular distances of individual galaxies.
One can, however, use this to measure the relative distances of two galaxy
samples provided each sample has enough galaxies to reduce the mean
intrinsic ellipticity of the sample well below the expected shear level.
Ideally one would choose the samples in some manner, such as by using
photometric redshifts, which would allow the galaxies inside each sample
to be at a similar distance.  It is, however, still
possible to measure a mean angular distance ratio for two sets of galaxies,
each of which has a broad redshift distribution.

\begin{figure}
\centering
\resizebox{\hsize}{!}{\includegraphics{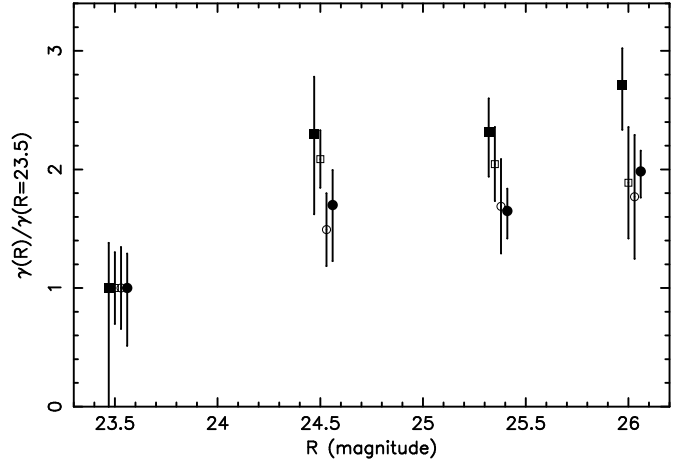}}
\caption{Shown above are values for the mean shear for the background galaxies,
divided into four magnitude bins (23-24, 24-25, 25-25.7, and 25.7-26.3),
relative to the mean shear of the brightest magnitude bin.  Only galaxies
located further than $350 h^{-1}$ kpc from the cluster centers were used
to compute the mean shear.
The mean shear of the three $z\sim 0.55-0.6$ clusters are given by the open
circles and the mean shear of the three $z\sim 0.8$ clusters are given
by the open squares.  The filled circles and squares are the expected shear
levels based on the photometric redshifts of the HDF-S galaxies in the same
magnitude bins for clusters at $z=0.55$ and $z=0.8$ respectively.}
\label{fig4}
\end{figure}

In the weak lensing limit, where $g \approx \gamma $, the shear acting upon a
galaxy is a function of the lens mass, the galaxy position, the lens and
galaxy redshifts, and the cosmological model.  If the galaxy samples
being compared have the same spatial distribution about a common lens, then
the ratio of the mean shears is a function only of the redshift of the lens,
the redshift distributions of the samples, and the cosmological model.  
If the magnification of the background galaxies is corrected for, the galaxy 
samples around different lenses of similar redshift can be coadded to improve 
the signal-to-noise of the mean shear ratio.

In Fig.~\ref{fig4} we show the relative strength of the mean shear signal for the 
three $z\sim 0.8$ clusters and the three $z\sim 0.55$ clusters in four
magnitude bins.  For both sets of clusters, the strength of the shear signal
increases with increasing magnitude, with significances, calculated from 
Student's t-distribution, of $96.4\%, 72.4\%,$ and $96.0\%$ for the 
$z\sim 0.5$ clusters, $z\sim 0.8$ clusters, and both sets combined 
respectively.  This is consistent with the mean redshift
of the background galaxies increasing with magnitude.  Also shown in
Fig.~\ref{fig4} are the shears which would be measured from the Fontana 
HDF-S photometric redshifts when using the same magnitude bins.  

Due to $\Sigma _{\mathrm{crit}}^{-1}$ increasing more rapidly for higher
redshift lenses, one should, in theory, be able to use multiple lenses at
different redshifts to obtain estimates for the redshift distribution of the
background galaxies.  This can be seen in Fig.~\ref{fig4} in which the difference
in the lensing strength predicted by the HDF-S photometric redshift catalogs 
for the $z\sim 0.8$ and $z\sim 0.55$ lenses continues to increase with
increasing magnitude of the background galaxies.  This difference, however,
is too small to measure with this data set.  We estimate that we would need
a data set ten times as large (60 clusters) with the same quality of data
in order to successfully apply any of the techniques \citep[e.g.][]{BA95.1} to measure the background galaxy redshift distribution.

\section{Discussion and Conclusions}
We have shown that a comparison of weak lensing shear profiles and X-ray
$\beta$-models result in best-fit values of $\Sigma _{\mathrm{crit}}^{-1}$
in good agreement with the photometric redshift distribution \citep{FO99.1} 
of the faint galaxies in the HDF-S.  This is true for both a
model-independent derivation of the weak lensing mass at a given radius and
for directly fitting the weak lensing mass with the X-ray $\beta$-model.
For both comparison methods, assuming the mean lensing redshift of the HDF-S
would result in the weak lensing mass measurements being higher than the X-ray
models for both of the $z\sim 0.5$ clusters, while the weak lensing masses
would be less than the X-ray mass model for all three of the $z\sim 0.8$
clusters.  For both sets of clusters, the difference is only of marginal
significance.  It should be noted, however, that the $\beta$-model fit
for \object{RXJ 1716.4$+$6708} does not provide a good fit to the ROSAT data
\citep{GI99.1}, and that the \object{MS 1054.4$-$0321} model does not include
an extended structure to the west of the cluster core \citep{NE00.1}
which is included in the weak lensing measurements.

One source of systematic error in the weak lensing mass estimates can be
the dilution of the shear signal from blue cluster dwarf galaxies.  The
background galaxy catalogs were selected from all detected galaxies with 
$23<R<26.3$ and $R-I<0.9$.  The color selection removed the red-sequence
of cluster ellipticals from the galaxy catalogs, but would have left some
fraction of the bluer cluster galaxies.  Cluster galaxies at $z\sim 0.8$
are redder in $R-I$ than those at $z\sim 0.5$.  As a result, applying the
same color cut to both sets of clusters would remove a greater fraction of
cluster spirals from the $z\sim 0.8$ background galaxy catalogs than from the
$z\sim 0.5$ catalogs.  From number counts of dwarf galaxies in nearby clusters
\citep[e.g.][]{TR98.2}, we estimate that the weak lensing shear signal, and thus
the derived masses, could be under-predicted by $10-20\%$ for the $z\sim 0.5$
clusters.  This estimate, however, depends greatly on a lack of evolution in
the number counts of dwarf galaxies compared to the cluster $L_*$ population.

We also compared the ratio of the shear signals as a function of magnitude,
and demonstrate that the measured shear does tend to increase with
increasing magnitude.  The amount of the increase is again in good agreement
with the photometric redshifts of the HDF-S.  This result is also in agreement
with that of \citet{HO00.1}, who compared the relative lensing strength
of galaxies in an HST mosaic of \object{MS 1054.4$-$0321}.  The level of noise in our
comparison, however, is too great to attempt to obtain a meaningful mean
lensing redshift as a function of magnitude for the background galaxies.

\begin{acknowledgements}
We thank Gillian Wilson, Lev Koffman, Len Cowie, Dave Sanders, 
John Learned, and Peter Schneider for their help and advice.  We also 
wish to thank Megan Donahue, Isabella Gioia, and J.~Patrick Henry for sharing
their X-ray data with us.  This work was supported
by NSF Grants AST-9529274 and AST-9500515, Nasa Grant NAG5-2594,
ASI-CNR, and the Deutsche Forschungsgemeinschaft under
the project SCHN 342/3--1.
\end{acknowledgements}

\bibliographystyle{apj}
\bibliography{H4502.bib}
\end{document}